\documentclass[aps,prl,twocolumn,groupedaddress,showpacs]{revtex4}

\usepackage{amssymb,amsmath,graphicx}

\begin{document}

\title{
Topological Defect Density in One-Dimensional \\
Friedmann--Robertson--Walker Cosmological Model:
Corrections \\
Inferred from the Multi-Josephson-Junction-Loop Experiment
}

\author{Yurii V. \surname{Dumin}}

\email[E-mail: ]{dumin@yahoo.com}

\affiliation{
IZMIRAN, Russian Academy of Sciences,
Troitsk, Moscow region, 142190 Russia
}

\date{19 August 2003}

\begin{abstract}
The data on a strongly-nonequilibrium superconducting phase transition in
the multi-Josephson-junction loop (MJJL), which represents a close analog
of one-dimensional Friedmann--Robertson--Walker cosmological model,
are used to refine a concentration of topological defects after
the phase transitions of Higgs fields in the early Universe.
The thermal correlations between the phases of order parameter
revealed in MJJL can reduce considerably the expected number of
cosmological defects and, thereby, show a new way to resolve
the long-standing problem of their excessive concentration.
\end{abstract}


%
%
%
%


\maketitle

\section{Introduction}

An important feature of the phase transitions of Higgs fields
expected at the early stages of cosmological evolution
is formation of topological defects,
whose concentration can be roughly estimated as
$
n \approx 1 / \, {\xi}_{\rm eff}^{\,d}.
$
Here, $ d = $~3, 2, and 1 for the monopoles, cosmic strings,
and domain walls, respectively; and
$ {\xi}_{\rm eff} $ is the effective correlation length,
which is commonly assumed to be less than
the cosmological causality horizon,
$
{\xi}_{\rm eff} \, \lesssim \, c \, / H_{\rm pt} \, ,
$
where $ c $ is the speed of light,
and $ H_{\rm pt} $ is Hubble constant at the instant of phase transition.
So, the resulting concentration of the defects should be
\begin{equation}
n \, \gtrsim \, \left( H_{\rm pt} / c \right){}^d \, ,
\label{standard_limit}
\end{equation}
which exceeds considerably the upper limits following from
the observational data (e.g.\ review~\cite{Klapdor97}).

For the sake of definiteness, we shall further consider
the simplest type of defects---the domain walls.
In such a case, their excessive concentration
can result in the $t^2$-dependence for the evolution of
cosmological scale factor, leaving less time for galaxy formation
and changing the rate of nucleosynthesis.
In addition, the domain walls can produce unreasonably large distortions
in the cosmic microwave background radiation~\cite{Zeldovich74,Gelmini89}.

A commonly-used approach to resolve the domain-wall problem
is to introduce some mechanism of their annihilation, for example,
utilizing the concept of the so-called ``biased'' (or asymmetric)
vacuum~\cite{Zeldovich74}. As a result,
under appropriate choice of parameters of the fields involved,
the regions of ``false'' (energetically unfavorable) vacuum
will quickly disappear, eliminating the corresponding domain walls.
(A detailed discussion of the various regimes of evolution
can be found in Ref.~\cite{Gelmini89}.)

Unfortunately, the concept of the biased vacuum is not
sufficiently supported by realistic models of elementary particles.
So, it becomes interesting to seek for solution of the problem
by another way, namely, to answer the question
if there are some mechanisms reducing the commonly-used lower limit
on the topological defect concentration~(\ref{standard_limit}).
To examine the efficiency of formation of the defects
by the strongly-nonequilibrium symmetry-breaking phase transitions,
a number of experiments was carried out recently
in the condensed-matter systems,
such as liquid crystals~\cite{Chuang91,Bowick94},
superfluid ${}^4\text{He}$~\cite{Dodd98}
and ${}^3\text{He}$~\cite{Bauerle96,Ruutu96},
and superconductors both in the form of bulk samples~\cite{Carmi99}
and quasi-one-dimensional structures~\cite{Carmi00}.
(The comprehensive review was given, for example,
by Kibble~\cite{Kibble01}.)

As was established in the last-mentioned experiment,
utilizing the multi-Josephson-junction loop (MJJL)~\cite{Carmi00},
the probability of occurrence of various field configurations
during the phase transition can be estimated much better
if the energy concentrated in the defects is taken into account.
The main aim of the present article is to apply this idea
to the consideration of cosmological phase transitions and, thereby,
to find the range of parameters at which formation of the domain walls
can be substantially suppressed even without any assumption of
asymmetry (or bias) of the vacuum states.

\section{Review of the MJJL Experiment}

\begin{figure}
\includegraphics[width=7cm]{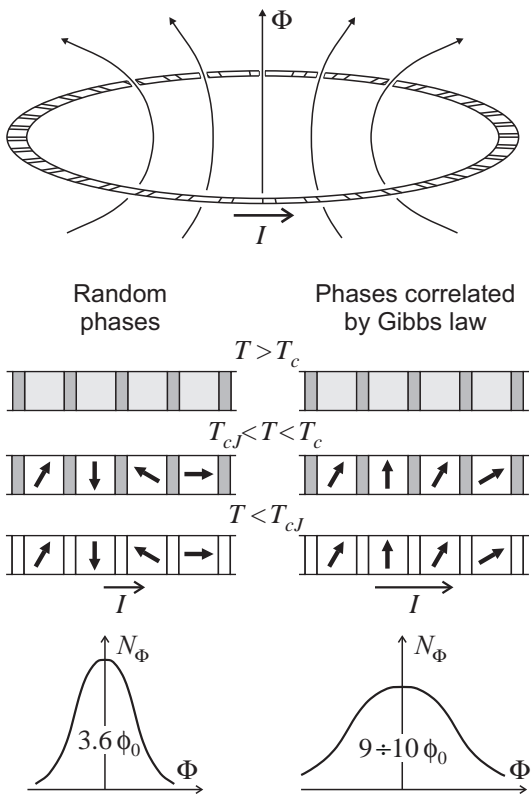}
\caption{\label{MJJL_exp}
Sketch of the MJJL experiment.
}
\end{figure}

A general scheme of the MJJL experiment is shown in Fig.~\ref{MJJL_exp}:
a thin loop produced of the high-temperature
YBa${}_2$Cu${}_3$O${}_{7\text{-}\delta}$ superconductor and
interrupted by 214 Josephson junctions at the grain boundaries
is rapidly cooled from the normal to superconducting phase
(namely, from ${\sim}100$~K to 77~K)
\footnote{
The actual loop used in the experiment is a winding strip
engraved at the boundary between grains of the superconductor film.
}.

At $T_c = 90$~K, the segments separated by junctions become
superconducting; however the junctions are still normal and,
therefore, the superconducting segments are effectively separate.
So, a random phase of the superconducting order parameter should
be established in each of them.

At subsequent cooling down to the temperature $T_{cJ}$
(which is $5{\div}7$~K below $T_c$),
the Josephson junctions also become superconducting,
so that a superconducting current $I$ is formed along the entire loop.
As a result, the loop will be penetrated by the magnetic flux $\Phi$,
which is just the measurable quantity.
(This is quite close to the original idea by Zurek~\cite{Zurek85},
who proposed to observe a spontaneous rotation produced by
the superfluid phase transition in a thin annular tube.)

Because of the phase jumps between the isolated segments
formed at the stage when $T_{cJ} < T < T_c$,
the final magnetic flux through the loop is nonzero and
varies randomly from one heating--cooling cycle to another.
Its histogram $N_{\Phi}(\Phi)$ derived from 166 cycles is well described
by the normal (Gaussian) law with zero average value and
standard deviation $7.4 \, {\phi}_0$
(where ${\phi}_0$ is the magnetic flux quantum)
\footnote{
The standard deviation of the total distribution is
just a typical value of the magnetic flux
measured in each individual heating--cooling cycle.
}.

In fact, the above-written experimental value is unreasonably large:
if the phase jumps between the segments
would be absolutely independent of each other,
then the expected width of the distribution should be only
$3.6 \, {\phi}_0$.
On the other hand, the excessive experimental value
was satisfactorily explained by the authors of this experiment
under assumption that the phases of superconducting order parameter
in the isolated segments are correlated to each other
so that probability $P({\delta}_i)$ of phase jump ${\delta}_i$
at the $i$'th junction is given by Gibbs law:
\begin{equation}
P({\delta}_i) \propto \exp [-{E_J}({\delta}_i) / k_B T] \, ,
\label{MJJL_corr}
\end{equation}
where
$E_J$ is the energy concentrated in the Josephson junction,
$T$ is the temperature, and
$k_B$ is Boltzmann constant.

So, the main conclusion from the results of the above experiment
is that the energy concentrated in the defects
should be taken into account in calculating the probability of
realization of the corresponding field configurations,
even if the defects are located at distances exceeding
the effective correlation length of the phase transition.
In other words, the correlation length represents
a size of the minimal domain in which the field must be uniform,
but it cannot be considered as the scale beyond which
the field states are absolutely independent of each other.
Of course, for the correlations at the larger scales to take place,
the corresponding regions should have the possibility
to interact with each other in the course of the previous evolution,
before the phase transition. (For the sake of brevity,
we shall call them to be in a coherent state.)
The last-mentioned condition is satisfied automatically
in the laboratory systems but requires a special consideration
in the cosmological applications
(see inequality~(\ref{coher_state}) below).

\section{Cosmological Implications}

First of all, we should emphasize close similarity between
the symmetry-breaking phase transitions in
the multi-Josephson-junction loop, drawn in Fig.~\ref{MJJL_exp}, and
the simplest one-dimensional (1D) Friedmann--Robertson--Walker (FRW)
cosmological model. To make the quantitative estimates,
let us consider the space--time metric
\begin{equation}
ds^2 = \: dt^2 - \, a^2(t) \: dx^2
\vphantom{X_X}
\label{metric}
\end{equation}
(where $a(t)$ is the scale factor of FRW model)
and the real scalar field $\varphi$ (simulating Higgs field)
whose Lagrangian
\begin{equation}
{\cal L} \, (x, t) \, = \,
\frac{1}{2} \, \big[ {\left( {\partial}_t \varphi \right)}^2 \! - \,
  {\left( {\partial}_x \varphi \right)}^2 \big]
  \, - \:
\frac{\lambda}{4} \,
  {\big[ \, {\varphi}^2 \! - \left( {\mu}^2 / \lambda \right) \big]}^2
\label{Lagrangian}
\end{equation}
possesses $\mathbb{Z}_2$ symmetry group, to be broken by the phase transition.
(The resulting topological defects in such a model will be evidently
the pointlike domain walls.)

As is known, the stable low-temperature vacuum states
of the field~(\ref{Lagrangian}) are
\begin{equation}
{\varphi}_0 = \, \pm \, \mu \, / \sqrt{\lambda} \: ,
\end{equation}
while a domain wall between them is described as
$
\varphi \, (x) = \, \pm \, {\varphi}_0
  \tanh \! \big[ ({\mu} / {\sqrt{2}}\,) \,
  ( x - x_0 ) \big]
$
and involves the energy
\begin{equation}
E = \, \frac{2 \, \sqrt{2}}{3} \, \frac{{\mu}^3}{\lambda} \; .
\label{d-w_energy}
\end{equation}
(From here on, it will be assumed that thickness of the wall
${\sim}{1 / \mu}$
is small in comparison with a characteristic domain size.)

Next, by introducing the conformal time
$ \eta = \! \int dt / a(t) \, $,
the space--time metric~(\ref{metric}) can be reduced to
the conformally flat form \cite{Misner69}:
\begin{equation}
ds^2 = a^2(t) \, [ \, d{\eta}^2 \! - dx^2 \, ] \, ;
\end{equation}
so that the light rays ($ ds^2 \! = 0 $)
will be described by the straight lines inclined at $ \pm \pi / 4 \, $:
$ x = \pm \eta + {\rm const} $.

\begin{figure}
\includegraphics[width=7.5cm]{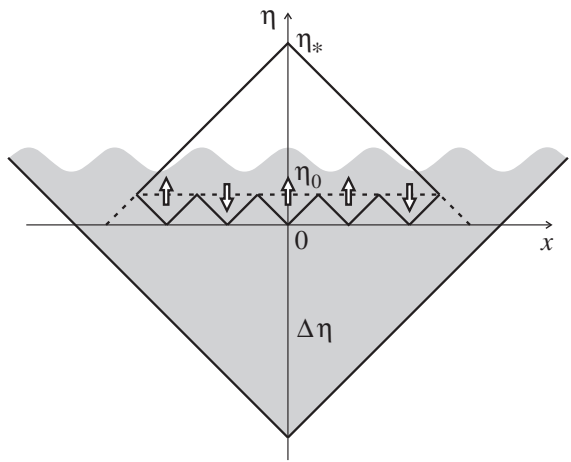}
\caption{\label{conf_diag}
Conformal diagram of the space--time.
}
\end{figure}

The entire structure of the space--time can be conveniently presented
by the conformal diagram in Fig.~\ref{conf_diag}.
Let $ \eta \! = \! 0 $ and $ \eta \! = \! {\eta}_0 $ be
the beginning and end of the phase transition, respectively,
and $ \eta \! = \! {\eta}_* $ be the instant of observation.
Then, as follows from a simple geometric consideration,
${\eta}_0$ is the maximum correlation length; so that
\begin{equation}
N \, = \, ( {\eta}_* \! - {\eta}_0 ) / {\eta}_0 \,
  \approx \, {\eta}_* / {\eta}_0
\quad (\text{at large } N)
\label{num_subreg}
\end{equation}
is the minimum number of spatial subregions causally-disconnected
\textit{during} the phase transition.
(Their final vacuum states are arbitrarily marked by the arrows.)

A probability of phase transition without formation of the domain walls
$P_N^{\,0}$ is commonly estimated as
ratio of the number of field configurations without domain walls
to their total number:
\begin{equation}
P_N^{\,0} \, = \, 2 \, / \, 2^N ,
\label{triv_probabil}
\end{equation}
which tends to zero very sharply at large $N$.
So, the observable part of the Universe,
represented by the large upper triangle in Fig.~\ref{conf_diag},
will inevitably contain a considerable number of the domain walls.

On the other hand, if a sufficiently long interval of the conformal time
\begin{equation}
\Delta \eta \, \ge \, {\eta}_*
\label{coher_state}
\end{equation}
preceded the phase transition,
then a coherent state of the Higgs field
(shown by the lower shaded triangle)
will be formed by the instant $ \eta \! = \! 0 $
in the entire region observable at $ \eta \! = \! {\eta}_* $.

The inequality~(\ref{coher_state}) can be satisfied, particularly,
in the case of sufficiently long de Sitter stage. Really, if
$ \, a(t) = \exp (Ht) $,
then
\begin{equation}
\eta \, = \, - \, \frac{1}{H} \, e^{-Ht} + {\rm const} \,
  \to \, - \infty \quad \text{at} \quad t \, \to \, - \infty \, ;
\end{equation}
so that $ \Delta \eta $ can be sufficiently large.
From the viewpoint of elementary-particle physics,
the de Sitter stage can be easily realized
in the overcooled state of Higgs field
just before its first-order phase transition.
(Let us remind that just this idea was the basis of
the first inflationary models;
for more details, see review~\cite{Linde84}.)

Next, if condition~(\ref{coher_state}) is satisfied,
it is reasonable to assume that the coherent state
of the field $\varphi$ will exhibit Gibbs-like correlations
(similar to the ones occurring in a superconducting Bose condensate
of MJJL) between the all $N$ subregions drawn in Fig.~\ref{conf_diag}.
In such a case, the probability $P_N^{\,0}$ should be calculated
taking into account the Gibbs factors for the field configurations
involving domain walls:
\begin{equation}
P_N^{\,0} \, = \, 2 \, / \, Z \: ,
\end{equation}
where
\begin{equation}
Z \, = \: \sum_{i=1}^N \;
  \sum_{s_i = \pm 1}
  \exp \bigg\lbrace \!
    - \frac{E}{T} \, \sum_{j=1}^N \,
    \frac{1}{2} \, (1 - \, s_{j} \, s_{j+1})
  \bigg\rbrace .
\label{stat_sum}
\end{equation}
Here,
$s_j$ is the spin-like variable describing a sign of the vacuum state
in the $j$'th subregion,
$E$ is the domain wall energy, given by~(\ref{d-w_energy}),
and $T$ is the characteristic temperature of the phase transition.

From the mathematical point of view,
statistical sum~(\ref{stat_sum}) is very similar to the one for Ising model,
well studied in the condensed-matter physics (e.g. Ref.~\cite{Isihara71}).
Using exactly the same methods
(attention should be paid to the appropriate choice of zero energy),
we get the final result:
\begin{equation}
P_N^{\,0} \, = \, \frac{2}{
  {\left[ 1 + e^{- E / T} \right]}^N +
  {\left[ 1 - e^{- E / T} \right]}^N
} \; .
\label{Gibbs_probabil}
\end{equation}
(Yet another method of calculating this quantity,
based on explicit expressions
for the probabilities of field configurations
with various numbers of the domain walls,
was presented in our article~\cite{Dumin00}.)

\begin{figure}
\includegraphics[width=7.5cm]{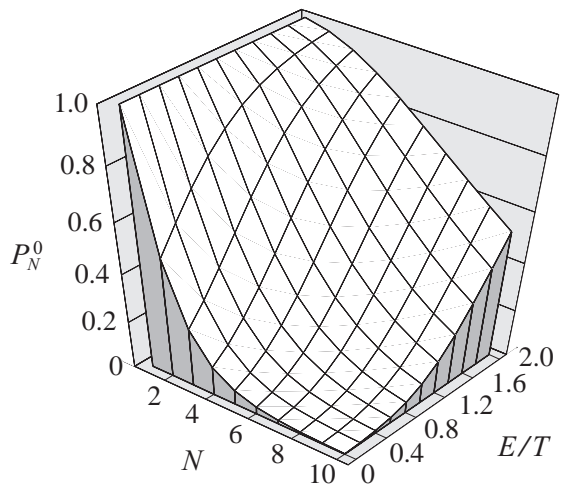}
\caption{\label{prob_distr}
The probability of phase transition
without formation of the domain walls~$P_N^{\,0}$
as function of the number of disconnected subregions~$N$
and the ratio of the domain wall energy to
the phase transition temperature~$E/T$.
}
\end{figure}

As is seen in Fig.~\ref{prob_distr},
$P_N^{\,0}$ drops very sharply with increasing $N$
at small values of $E/T$
\footnote{
$E/T = 0$ corresponds to the case when
Gibbs-like correlations between the subregions are absent at all,
so that formula~(\ref{Gibbs_probabil}) coincides
with~(\ref{triv_probabil}).
}
but becomes a gently decreasing function of $N$ when
the parameter $E/T$ is sufficiently large. Therefore,
\textit{just the large energy concentrated in the domain walls
turns out to be the factor substantially suppressing
the probability of their formation.}

As can be easily derived from~(\ref{Gibbs_probabil}),
the probability of absence of the domain walls
in the observable region of space--time
becomes on the order of unity (for example, $1/2$) if
$\, E/T \gtrsim \ln N $
or, by substituting (\ref{d-w_energy}) and (\ref{num_subreg}),
\begin{equation}
\frac{{\mu}^3}{\lambda \, T} \: \gtrsim \:
  \ln \frac{{\eta}_*}{{\eta}_0} \: ,
\label{d-w_absence}
\end{equation}
which is just the required inequality, mentioned in Introduction.
Because of the very weak logarithmic dependence
in the right-hand side of~(\ref{d-w_absence}),
this condition may be quite reasonable
for a certain class of field theories.

Moreover, the situation should be even more favorable
in the cases of higher spatial dimensions.
A well-known property of Ising models in 2 and 3 dimensions
is the tendency for aggregation of the domains
with the same sign of the order parameter
when the temperature drops below some critical value
(e.g. Ref.~\cite{Rumer80}).
In the condensed-matter applications,
this corresponds to the spontaneous magnetization of a solid body.
Regarding the cosmological context, we can expect that
probability of formation of the domain walls will be reduced dramatically
at the sufficiently large values of $E/T$
\footnote{
To avoid misunderstanding, it should be emphasized once again
that the Ising model considered here
is only an auxiliary mathematical construction,
describing a final distribution of the domain walls
after the phase transition.
This physical phase transition of the field $\varphi$
bears no relation to the phase transition in Ising model.
}.
This subject will be discussed in more detail in
a separate paper~\cite{Dumin03b}.

Although the model under consideration requires a presence of
inflationary stage (for the coherent state of Higgs field to be formed
\footnote{
Besides, the inequality~(\ref{d-w_absence}) can be satisfied
much more easily if the phase transition develops from
a strongly overcooled state of the Higgs field.
}),
it is not equivalent to the traditional inflationary models,
where decrease in concentration of
the \textit{previously formed} topological defects
is a purely geometrical effect,
associated with sharp expansion of the space.
This raises a problem of \textit{new} topological defects
that can be formed at a \textit{subsequent stage}
of decay of the inflaton field by itself.
On the other hand, the inflationary stage in our model
results in \textit{suppressing formation} of the defects
(rather than subsequent decrease in their concentration).
From this point of view,
it represents a self-consistent solution of the problem.

\begin{acknowledgments}

This work was partially supported by
the ESF COSLAB
Programme and
the Abdus Salam ICTP.
I am grateful to
V.B.~Belyaev,
A.M.~Che\-chel\-nitsky,
I.~Coleman,
V.B.~Efimov,
H.J.~Junes,
I.B.~Khrip\-lo\-vich,
T.W.B.~Kibble,
M.~Knyazev,
O.D.~Lav\-ren\-to\-vich,
V.N.~Lukash,
P.V.E.~McClintock,
L.B.~Okun,
E.~Polturak,
A.I.~Rez,
M.~Sasaki,
A.Yu.~Smirnov,
A.A.~Sta\-ro\-binsky,
A.V.~To\-po\-rensky,
W.G.~Unruh,
G.E.~Volo\-vik, and
W.H.~Zu\-rek
for valuable discussions, consultations, and critical comments.

\end{acknowledgments}



\begin{thebibliography}{18}
\expandafter\ifx\csname natexlab\endcsname\relax\def\natexlab#1{#1}\fi
\expandafter\ifx\csname bibnamefont\endcsname\relax
  \def\bibnamefont#1{#1}\fi
\expandafter\ifx\csname bibfnamefont\endcsname\relax
  \def\bibfnamefont#1{#1}\fi
\expandafter\ifx\csname citenamefont\endcsname\relax
  \def\citenamefont#1{#1}\fi
\expandafter\ifx\csname url\endcsname\relax
  \def\url#1{\texttt{#1}}\fi
\expandafter\ifx\csname urlprefix\endcsname\relax\def\urlprefix{URL }\fi
\providecommand{\bibinfo}[2]{#2}
\providecommand{\eprint}[2][]{\url{#2}}

\bibitem[{\citenamefont{Klapdor-Kleingrothaus and Zuber}(1997)}]{Klapdor97}
\bibinfo{author}{\bibfnamefont{H.}~\bibnamefont{Klapdor-Kleingrothaus}}
  \bibnamefont{and} \bibinfo{author}{\bibfnamefont{K.}~\bibnamefont{Zuber}},
  \emph{\bibinfo{title}{Particle Astrophysics}} (\bibinfo{publisher}{Inst.
  Phys. Publ., Bristol}, \bibinfo{year}{1997}).

\bibitem[{\citenamefont{Zel'dovich et~al.}(1974)\citenamefont{Zel'dovich,
  Kobzarev, and Okun}}]{Zeldovich74}
\bibinfo{author}{\bibfnamefont{Ya.}~\bibnamefont{Zel'dovich}},
  \bibinfo{author}{\bibfnamefont{I.}~\bibnamefont{Kobzarev}}, \bibnamefont{and}
  \bibinfo{author}{\bibfnamefont{L.}~\bibnamefont{Okun}},
  \bibinfo{journal}{Zh.\ Eksp.\ Teor.\ Fiz.} \textbf{\bibinfo{volume}{67}},
  \bibinfo{pages}{3} (\bibinfo{year}{1974}) \bibinfo{note}{[Sov.\ Phys.---JETP
  \textbf{40}, 1 (1975)]}.

\bibitem[{\citenamefont{Gelmini et~al.}(1989)\citenamefont{Gelmini, Gleiser,
  and Kolb}}]{Gelmini89}
\bibinfo{author}{\bibfnamefont{G.}~\bibnamefont{Gelmini}},
  \bibinfo{author}{\bibfnamefont{M.}~\bibnamefont{Gleiser}}, \bibnamefont{and}
  \bibinfo{author}{\bibfnamefont{E.}~\bibnamefont{Kolb}},
  \bibinfo{journal}{Phys.\ Rev.\ D} \textbf{\bibinfo{volume}{39}},
  \bibinfo{pages}{1558} (\bibinfo{year}{1989}).

\bibitem[{\citenamefont{Chuang et~al.}(1991)\citenamefont{Chuang, Durrer,
  Turok, and Yurke}}]{Chuang91}
\bibinfo{author}{\bibfnamefont{I.}~\bibnamefont{Chuang}},
  \bibinfo{author}{\bibfnamefont{R.}~\bibnamefont{Durrer}},
  \bibinfo{author}{\bibfnamefont{N.}~\bibnamefont{Turok}}, \bibnamefont{and}
  \bibinfo{author}{\bibfnamefont{B.}~\bibnamefont{Yurke}},
  \bibinfo{journal}{Science} \textbf{\bibinfo{volume}{251}},
  \bibinfo{pages}{1336} (\bibinfo{year}{1991}).

\bibitem[{\citenamefont{Bowick et~al.}(1994)\citenamefont{Bowick, Chandar,
  Shiff, and Srivastava}}]{Bowick94}
\bibinfo{author}{\bibfnamefont{M.}~\bibnamefont{Bowick}},
  \bibinfo{author}{\bibfnamefont{L.}~\bibnamefont{Chandar}},
  \bibinfo{author}{\bibfnamefont{E.}~\bibnamefont{Shiff}}, \bibnamefont{and}
  \bibinfo{author}{\bibfnamefont{A.}~\bibnamefont{Srivastava}},
  \bibinfo{journal}{Science} \textbf{\bibinfo{volume}{263}},
  \bibinfo{pages}{943} (\bibinfo{year}{1994}).

\bibitem[{\citenamefont{Dodd et~al.}(1998)\citenamefont{Dodd, Hendry, Lawson,
  McClintock, and Williams}}]{Dodd98}
\bibinfo{author}{\bibfnamefont{M.}~\bibnamefont{Dodd}},
  \bibinfo{author}{\bibfnamefont{P.}~\bibnamefont{Hendry}},
  \bibinfo{author}{\bibfnamefont{N.}~\bibnamefont{Lawson}},
  \bibinfo{author}{\bibfnamefont{P.}~\bibnamefont{McClintock}},
  \bibnamefont{and} \bibinfo{author}{\bibfnamefont{C.}~\bibnamefont{Williams}},
  \bibinfo{journal}{Phys.\ Rev.\ Lett.} \textbf{\bibinfo{volume}{81}},
  \bibinfo{pages}{3703} (\bibinfo{year}{1998}).

\bibitem[{\citenamefont{B{\"a}uerle et~al.}(1996)\citenamefont{B{\"a}uerle,
  Bunkov, Fisher, Godfrin, and Pickett}}]{Bauerle96}
\bibinfo{author}{\bibfnamefont{C.}~\bibnamefont{B{\"a}uerle}},
  \bibinfo{author}{\bibfnamefont{Yu.}~\bibnamefont{Bunkov}},
  \bibinfo{author}{\bibfnamefont{S.}~\bibnamefont{Fisher}},
  \bibinfo{author}{\bibfnamefont{H.}~\bibnamefont{Godfrin}}, \bibnamefont{and}
  \bibinfo{author}{\bibfnamefont{G.}~\bibnamefont{Pickett}},
  \bibinfo{journal}{Nature (London)} \textbf{\bibinfo{volume}{382}},
  \bibinfo{pages}{332} (\bibinfo{year}{1996}).

\bibitem[{\citenamefont{Ruutu et~al.}(1996)\citenamefont{Ruutu, Eltsov, Gill,
  Kibble, Krusius, Makhlin, Pla{\c{c}}ais, Volovik, and {Wen Xu}}}]{Ruutu96}
\bibinfo{author}{\bibfnamefont{V.}~\bibnamefont{Ruutu}},
  \bibinfo{author}{\bibfnamefont{V.}~\bibnamefont{Eltsov}},
  \bibinfo{author}{\bibfnamefont{A.}~\bibnamefont{Gill}},
  \bibinfo{author}{\bibfnamefont{T.}~\bibnamefont{Kibble}},
  \bibinfo{author}{\bibfnamefont{M.}~\bibnamefont{Krusius}},
  \bibinfo{author}{\bibfnamefont{Yu.}~\bibnamefont{Makhlin}},
  \bibinfo{author}{\bibfnamefont{B.}~\bibnamefont{Pla{\c{c}}ais}},
  \bibinfo{author}{\bibfnamefont{G.}~\bibnamefont{Volovik}}, \bibnamefont{and}
  \bibinfo{author}{\bibnamefont{{Wen Xu}}}, \bibinfo{journal}{Nature (London)}
  \textbf{\bibinfo{volume}{382}}, \bibinfo{pages}{334} (\bibinfo{year}{1996}).

\bibitem[{\citenamefont{Carmi and Polturak}(1999)}]{Carmi99}
\bibinfo{author}{\bibfnamefont{R.}~\bibnamefont{Carmi}} \bibnamefont{and}
  \bibinfo{author}{\bibfnamefont{E.}~\bibnamefont{Polturak}},
  \bibinfo{journal}{Phys.\ Rev.\ B} \textbf{\bibinfo{volume}{60}},
  \bibinfo{pages}{7595} (\bibinfo{year}{1999}).

\bibitem[{\citenamefont{Carmi et~al.}(2000)\citenamefont{Carmi, Polturak, and
  Koren}}]{Carmi00}
\bibinfo{author}{\bibfnamefont{R.}~\bibnamefont{Carmi}},
  \bibinfo{author}{\bibfnamefont{E.}~\bibnamefont{Polturak}}, \bibnamefont{and}
  \bibinfo{author}{\bibfnamefont{G.}~\bibnamefont{Koren}},
  \bibinfo{journal}{Phys.\ Rev.\ Lett.} \textbf{\bibinfo{volume}{84}},
  \bibinfo{pages}{4966} (\bibinfo{year}{2000}).

\bibitem[{\citenamefont{Kibble}(2001)}]{Kibble01}
\bibinfo{author}{\bibfnamefont{T.}~\bibnamefont{Kibble}},
  \emph{\bibinfo{title}{Testing cosmological defect formation in the
  laboratory}} (\bibinfo{year}{2001}), \eprint{cond-mat/0111082}.

\bibitem[{\citenamefont{Zurek}(1985)}]{Zurek85}
\bibinfo{author}{\bibfnamefont{W.}~\bibnamefont{Zurek}},
  \bibinfo{journal}{Nature (London)} \textbf{\bibinfo{volume}{317}},
  \bibinfo{pages}{505} (\bibinfo{year}{1985}).

\bibitem[{\citenamefont{Misner}(1969)}]{Misner69}
\bibinfo{author}{\bibfnamefont{C.}~\bibnamefont{Misner}},
  \bibinfo{journal}{Phys.\ Rev.\ Lett.} \textbf{\bibinfo{volume}{22}},
  \bibinfo{pages}{1071} (\bibinfo{year}{1969}).

\bibitem[{\citenamefont{Linde}(1984)}]{Linde84}
\bibinfo{author}{\bibfnamefont{A.}~\bibnamefont{Linde}},
  \bibinfo{journal}{Rep.\ Progr.\ Phys.} \textbf{\bibinfo{volume}{47}},
  \bibinfo{pages}{925} (\bibinfo{year}{1984}).

\bibitem[{\citenamefont{Isihara}(1971)}]{Isihara71}
\bibinfo{author}{\bibfnamefont{A.}~\bibnamefont{Isihara}},
  \emph{\bibinfo{title}{Statistical Physics}} (\bibinfo{publisher}{Academic
  Press, NY}, \bibinfo{year}{1971}).

\bibitem[{\citenamefont{Dumin}(2000)}]{Dumin00}
\bibinfo{author}{\bibfnamefont{Yu.}~\bibnamefont{Dumin}}, in
  \emph{\bibinfo{booktitle}{Proc.\ Int.\ Workshop Hot Points in Astrophysics}}
  (\bibinfo{publisher}{JINR, Dubna}, \bibinfo{year}{2000}), p.
  \bibinfo{pages}{114}.

\bibitem[{\citenamefont{Rumer and Ryvkin}(1980)}]{Rumer80}
\bibinfo{author}{\bibfnamefont{Yu.}~\bibnamefont{Rumer}} \bibnamefont{and}
  \bibinfo{author}{\bibfnamefont{M.}~\bibnamefont{Ryvkin}},
  \emph{\bibinfo{title}{Thermodynamics, Statistical Physics, and Kinetics}}
  (\bibinfo{publisher}{Mir, Moscow}, \bibinfo{year}{1980}).

\bibitem[{\citenamefont{Dumin and Svirskaya}(2003)}]{Dumin03b}
\bibinfo{author}{\bibfnamefont{Yu.}~\bibnamefont{Dumin}} \bibnamefont{and}
  \bibinfo{author}{\bibfnamefont{L.}~\bibnamefont{Svirskaya}},
  \emph{\bibinfo{title}{On the efficiency of defect formation in the systems of
  various size and (quasi-) dimensionality}} (\bibinfo{year}{2003}),
  \bibinfo{note}{in press}.

\end{thebibliography}
\end{document}